\documentclass[a4paper, 12pt, leqno]{article}
\usepackage[utf8]{inputenc}
\usepackage{indentfirst}
\usepackage{fullpage}
\usepackage{amsmath}
\usepackage{amsfonts}
\usepackage{amssymb}
\usepackage{amsthm}
\usepackage{enumitem}
\usepackage[vlined]{algorithm2e}
\usepackage{multicol}
\usepackage{graphicx}

\numberwithin{equation}{section}
\DontPrintSemicolon
\newtheorem{theorem}{Theorem}[section]
\newtheorem{lemma}[theorem]{Lemma}
\setlist{itemsep=0pt, topsep=0pt}

\newcommand\s{\operatorname{s}}
\newcommand\mul{\operatorname{Mul}}
\newcommand\T{\operatorname{T}}

\begin{document}

\setlength\abovedisplayshortskip{0.2ex}
\setlength\belowdisplayshortskip{0.2ex}
\setlength\abovedisplayskip{0.4ex}
\setlength\belowdisplayskip{0.4ex}

\title{\bfseries Algebraic number fields and the LLL algorithm}

\author{
	M.\ János Uray \\
	uray.janos@inf.elte.hu \\
	ELTE -- Eötvös Loránd University (Budapest) \\
	Faculty of Informatics \\
	Department of Computer Algebra
}
\date{}

\maketitle

\begin{abstract}
In this paper we analyze the computational costs of various operations and algorithms in algebraic number fields using exact arithmetic. Let $K$ be an algebraic number field. In the first half of the paper, we calculate the running time and the size of the output of many operations in $K$ in terms of the size of the input and the parameters of $K$. We include some earlier results about these, but we go further than them, e.g.\ we also analyze some $\mathbb{R}$-specific operations in $K$ like less-than comparison.

In the second half of the paper, we analyze two algorithms: the Bareiss algorithm, which is an integer-preserving version of the Gaussian elimination, and the LLL algorithm, which is for lattice basis reduction. In both cases, we extend the algorithm from $\mathbb{Z}^n$ to $K^n$, and give a polynomial upper bound on the running time when the computations in $K$ are performed exactly (as opposed to floating-point approximations).
\end{abstract}

\section{Introduction} \label{sec:intro}

Exact computation with algebraic numbers is an important feature that most computer algebra systems provide. They use efficient algorithms for the calculations, described in several papers and books, e.g.\ in \cite{cohen, geddes, pohst-zassenhaus}. However, the computational costs of these algorithms are often not obvious to calculate, because the bit complexity depends on how much the representing multi-precision integers grow during the computation.

In this paper, given an algebraic number field $K$, we give explicit bounds on the costs of many operations in $K$, using the size of the input and the parameters of $K$. We consider polynomial representation, i.e.\ when the elements are represented by polynomials of a fixed primitive element $\theta \in K$. There are already some work in this area, e.g.\ \cite{belabas} and \cite{comp-hnf}, and we use the existing results whenever we can, but in some cases, we replace them with more general or in other ways more suitable results. We also consider operations that are special to $\mathbb{R}$, like less-than comparison and integer rounding functions, and the author is not aware of any previous results on these.

The obtained explicit formulas enable us to calculate the running time of several well-known algorithms if we apply them to $K$ with exact arithmetic. For example consider the Gaussian elimination, which performs $O(n^3)$ arithmetic operations, but if we use exact arithmetic in $\mathbb{Z}$ or $K$, then the size of the entries may grow exponentially. The Bareiss algorithm \cite{bareiss} is a modification for $\mathbb{Z}$ which addresses this problem by certain simplifications to ensure polynomial coefficient growth and running time (although with larger exponent). The idea can be generalized to $K$ in a straightforward manner, which is calculated in this paper.

Another algorithm of our interest is the LLL lattice reduction algorithm \cite{lll}, which converts a basis of a lattice to a reduced basis of the same lattice. It is a well-studied algorithm for integers and floating-point numbers. The creators of the algorithm showed in \cite{lll} that the algorithm runs in polynomial time if the input vectors are in $\mathbb{Z}^n$. There are also many efficient generalizations for $\mathbb{R}^n$ using floating-point arithmetic (\cite{l2-alg, h-lll, lll-float-adaptive, lll-msb}). But we are interested in another generalization: we consider the LLL algorithm on vectors in $K^n$ for real $K$ with exact arithmetic. The LLL algorithm over number fields has already been studied (e.g.\ in \cite{lll-short-bases, lll-module}), but we are not aware of a complexity analysis using exact arithmetic. In the present paper, using the calculated operation costs in $K$, we analyse the algorithm: we find an upper bound on the number of main iterations, examine how the sizes of the variables grow during these iterations, and conclude with an explicit upper bound on the running time of the algorithm. All these results are given in terms of very basic parameters like the dimension, the properties of $K$ and the bit size of the input. The running time turns out to be polynomial in these parameters, though with larger exponents than in $\mathbb{Z}$.

Exact calculation in the LLL algorithm might not seem very useful, since for many practical applications, the goal is to find a well-reduced basis or a sufficiently short vector, therefore the floating-point LLL versions can be applied, and they are much faster than exact computation. However, we still think that the analysis of the exact algorithm in $K^n$ deserves interest. First, it is interesting from a theoretical point of view. Second, there are applications when the exact values in the reduction are needed. For example in \cite{petho-lll}, algebraic integers are represented by ultimately periodic series of integer vectors, obtained by a repeated application of the LLL algorithm. This representation is a generalization of continued fractions, and as with continued fractions, the exact representation is not guaranteed to be obtained if we use approximations. This is analoguous to the Gaussian elimination where the floating-point version is preferred whenever applicable, but if we need the exact result, we have no choice but to use a slower but exact version like the Bareiss algorithm.

The paper is built up as follows: Section~\ref{sec:alg} analyzes the computational costs of several operations in $K$, Section~\ref{sec:bareiss} calculates the running time of the Bareiss algorithm over $K$, and Section~\ref{sec:lll} does the same with the LLL algorithm.

\section{Operations in number fields} \label{sec:alg}

Let $K$ be an algebraic number field of degree $m$, and let $\theta \in K$ be a primitive element, i.e.\ $K = \mathbb{Q}(\theta)$. For some results we will require that $\theta \in \mathbb{R}$ and so $K \subset \mathbb{R}$, but for the others, we state them more generally. Without loss of generality we can assume that $\theta$ is an algebraic integer. Denote its minimal polynomial by $f(x) = x^m + f_{m-1} x^{m-1} + \ldots + f_1 x + f_0\;(f_i \in \mathbb{Z}$). We will consider $f$, $\theta$ and $m$ as fixed throughout this article.

Elements of $K$ can be represented by rational linear combinations of $1, \theta, \theta^2, \ldots, \theta^{m-1}$. However, in order to minimalize the problems with rational numbers like simplification, we use an integer linear combination and a common denominator. Furthermore, we consider only the numerator, i.e.\ the ring $\mathbb{Z}[\theta]$, because dealing with the single denominator is trivial, and in many algorithms, they can be cleared in the beginning.

We define $H(f)$, the height of $f$, and a related quantity:
\begin{align}
	\label{def:Hf} H(f) &:= \max_{i=0}^{m-1} |f_i|, \\
	\label{def:F} F &:= \log(H(f) + 1).
\end{align}

(Note that in this paper, the bases of the logarithms are mostly omitted, but we mean the same fixed base throughout the paper. We can think of them as binary logarithms, because they mainly mean the number of bits, but the calculations remain consistent for any fixed $\geq 2$ base, as changing the base makes only a constant factor difference.)

In many results, the dependence on the number field will be expressed by the two constants $m$ and $F$. We start with two simple bounds on $\theta$:

\begin{lemma} \label{lem:theta-bound}
\begin{align}
	\label{theta-bound} \log |\theta| &< F, \\
	\label{theta-geom-bound} \log\left( 1 + |\theta| + |\theta|^2 + \ldots + |\theta|^{m-1} \right) &< m F.
\end{align}
\end{lemma}

\begin{proof}
Cauchy's inequality states that $|\theta| < H(f) + 1$ (see e.g.\ \cite[p.~323]{geddes}), which is equivalent to the first statement. The second one is a consequence:
\begin{align*}
	1 + |\theta| + |\theta|^2 + \ldots + |\theta|^{m-1}
	&< 1 + (H(f)+1) + (H(f)+1)^2 + \ldots + (H(f)+1)^{m-1} = \\
	&= \frac{(H(f)+1)^m - 1}{(H(f)+1) - 1}
	< (H(f)+1)^m.
\end{align*}
\end{proof}

\subsection{Size of the elements} \label{sec:alg-size}

We use some results from \cite[Sect.~3]{comp-hnf}. The authors use a more general representation: they fix an integral basis in the number field: $\Omega = \{\omega_1, \omega_2, \ldots, \omega_m\}$ with $\omega_1 = 1$, and write algebraic integers as $a_1 \omega_1 + a_2 \omega_2 + \ldots + a_m \omega_m$ for $a_1, \ldots, a_m \in \mathbb{Z}$.

They introduce three constants, $C_1$, $C_2$ and $C_3$, dependent on $\Omega$, and state their results in terms of these constants. First, they define $C_1$ and $C_2$ by stating in \cite[(1)]{comp-hnf} that (using our notation):
\begin{lemma} \label{lem:C1C2-exist}
Let $x := (a_0, a_1, \ldots, a_{m-1})^T$ be the coefficient vector of $a$, denote the roots of $f$ by $\theta_1, \theta_2, \ldots, \theta_m$, let $a^{(i)} := a_0 + a_1 \theta_i + \ldots + a_{m-1} \theta_i^{m-1}$, and let $y := (a^{(1)}, a^{(2)}, \ldots, a^{(m)})^T$. Then there exists such $C_1, C_2 > 0$ that
\begin{equation*}
	\frac{1}{C_2} \lVert x \rVert_\infty \leq \lVert y \rVert_2 \leq C_1 \lVert x \rVert_\infty.
\end{equation*}
\end{lemma}
Then they define $C_3$ to be a bound on the cofficients $m_{i,j,k}$ appearing in the multiplication table of $(\omega_i \omega_j)_{i,j}$:
\begin{gather*}
	\omega_i \omega_j = m_{i,j,1} \omega_1 + m_{i,j,2} \omega_2 + \ldots + m_{i,j,m} \omega_m, \\
	C_3 := \max_{i,j,k} |m_{i,j,k}|.
\end{gather*}

In order to use their results, we need to express $C_1$, $C_2$ and $C_3$ for our special case, $\Omega = \{1, \theta, \theta^2, \ldots, \theta^{m-1}\}$. We have the following result:

\begin{lemma} \label{lem:C1C2C3}
\begin{align}
	\label{C1} \log C_1 &< m F + \frac{1}{2} \log m, \\
	\label{C2} \log C_2 &< \binom{m}{2} F + m \log m, \\
	\label{C3} \log C_3 &< (m-1) F.
\end{align}
\end{lemma}

\begin{proof}
We can describe the connection between $x$ and $y$ from Lemma~\ref{lem:C1C2-exist} as $y = V x$ where $V$ is the following Vandermonde matrix:
\begin{equation*}
	V = \begin{pmatrix}
		1 & \theta_1 & \theta_1^2 & \ldots & \theta_1^{m-1} \\
		1 & \theta_2 & \theta_2^2 & \ldots & \theta_2^{m-1} \\
		\vdots & \vdots &  &  & \vdots \\
		1 & \theta_m & \theta_m^2 & \ldots & \theta_m^{m-1}
	\end{pmatrix}.
\end{equation*}
Since $\lVert y \rVert_\infty \leq \lVert y \rVert_2 \leq \!\sqrt{m} \,\lVert y \rVert_\infty$, we can choose $C_1$ and $C_2$ as
\begin{equation*}
	C_1 := \sqrt{m} \,\lVert V \rVert_\infty, \qquad
	C_2 := \lVert V^{-1} \rVert_\infty,
\end{equation*}
so we need to find $\lVert V \rVert_\infty$ and $\lVert V^{-1} \rVert_\infty$.

The former can be calculated using (\ref{theta-geom-bound}), thus proving (\ref{C1}):
\begin{equation*}
	\log \,\lVert V \rVert_\infty
	= \log \max_{i=1}^{m} (1 + |\theta_i| + |\theta_i|^2 + \ldots + |\theta_i|^{m-1})
	< m F.
\end{equation*}

For (\ref{C2}), we need $\lVert V^{-1} \rVert_\infty$. The inverse of $V$ can be written as $(V^{-1})_{ij} = C_{ji} / \det V$, where $C_{ij}$ is the minor of $V$ at row $i$ and column $j$ multiplied by $(-1)^{i+j}$. It is well-known that the determinant of the Vandermonde matrix is (see e.g.\ \cite[p.~185]{geddes}):
\begin{equation*}
	\det V = \prod_{i=1}^{m} \prod_{j=i+1}^{m} (\theta_j - \theta_i),
\end{equation*}
and also that its square is the discriminant of $f$ (see e.g.\ \cite[Prop.~3.3.5.]{cohen}). Since the latter is integer and nonzero, $|\det V| \geq 1$, so $|(V^{-1})_{ij}| \leq |C_{ji}|$. The cofactor $C_{ji}$ is an $(m-1)\times(m-1)$ determinant, whose full expansion has $(m-1)!$ terms, where each term is a product of $m-1$ elements of $V$ in different columns. Since $V_{kl} = \theta_k^{l-1}$, $\log |V_{kl}| < (l-1) F$ by (\ref{theta-bound}), and the sum of exponents in each product is at most $0 + 1 + 2 + \ldots + (m-1) = \binom{m}{2}$, we get $\log |C_{ji}| < \binom{m}{2} F + \log (m-1)!$. Therefore, we can finish (\ref{C2}) by:
\begin{align*}
	\log C_2 = \log \,\lVert V^{-1} \rVert_\infty
	&= \log \max_{i=1}^{m} \left( \left|(V^{-1})_{i1}\right| + \left|(V^{-1})_{i2}\right| + \ldots + \left|(V^{-1})_{im}\right| \right) \leq \\
	&\leq \log \max_{i=1}^{m} (|C_{1i}| + |C_{2i}| + \ldots + |C_{mi}|) < \\
	&< \binom{m}{2} F + \log (m-1)! + \log m
	\leq \binom{m}{2} F + m \log m.
\end{align*}

For (\ref{C3}), we write $\theta^{m+k}$ for $k = 0, 1, \ldots, m-2$ in terms of lower powers of $\theta$:
\begin{equation} \label{r-def}
	\theta^{m+k} = r_{k,0} + r_{k,1}\theta + r_{k,2}\theta^2 + \ldots + r_{k,m-1}\theta^{m-1},
\end{equation}
and we have $C_3 = \max_{k=0}^{m-2} \max_{j=0}^{m-1} |r_{k,j}|$.

By using that $f(\theta) = 0$, one can get a recursive formula for $r_{k,l}$ (see e.g.\ \cite[p.~159]{cohen}):
\begin{equation} \label{r-rec}
	r_{0,l} = -f_l, \qquad
	r_{k+1,l} = r_{k,l-1} - f_l r_{k,m-1},
\end{equation}
where $r_{k,-1} = 0$. Then one can easily show by induction that:
\begin{equation} \label{r-bound}
	|r_{k,l}| < (H(f)+1)^{k+1},
\end{equation}
and (\ref{C3}) follows.
\end{proof}

For an algebraic integer $a \in \mathbb{Z}[\theta]$, $a = a_0 + a_1\theta + a_2\theta^2 + \ldots + a_{m-1}\theta^{m-1}$ ($a_i \in \mathbb{Z}$), we will use the following function to measure its coefficient size:
\begin{equation}
	\s(a) := \begin{cases}
		\log \max_{i=0}^{m-1} |a_i| & \text{if }a \neq 0, \\
		0 & \text{if }a = 0.
	\end{cases}
\end{equation}
This quantity together with the field degree $m$ (which is constant for a fixed field) indicates the storage size needed by the algebraic integer $a$. (Note that when comparing results with \cite{comp-hnf}, their notation $\operatorname{S}(a)$ is slightly different: $\operatorname{S}(a) = m \s(a)$.) For an integer $n$, we simply have $\s(n) = \log |n|$ if $n \neq 0$, otherwise $0$. For convenience, we also define $\s(a_1, a_2, \ldots, a_n) := \max_{i=1}^{n} \s(a_i)$.

The following several results show how much $\s(\cdot)$ can grow during several operations. First we start with the most trivial ones.

\begin{lemma} \label{lem:triv-size}
\begin{align}
	\label{add-size}
		&a, b \in \mathbb{Z}[\theta]: &
		\s(a \pm b) &\leq \s(a, b) + \log 2, \\
	\label{addn-size}
		&n \in \mathbb{Z}^+\!,\; a_1, \ldots, a_n \in \mathbb{Z}[\theta]: &
		\s\left( a_1 + \ldots + a_n \right) &\leq \s(a_1, \ldots, a_n) + \log n, \\
	\label{intmul-size}
		&n \in \mathbb{Z},\; a \in \mathbb{Z}[\theta]: &
		\s(n a) &\leq \s(a) + \s(n).
\end{align}
\end{lemma}

For the multiplicative operations, \cite[before~Prop.~4]{comp-hnf} shows the following:
\begin{lemma} \label{lem:mult-size-C}
If $a, b \in \mathbb{Z}[\theta]$, then:
\begin{equation*}
	\s(a b) \leq \s(a) + \s(b) + \log C_3 + 2 \log m,
\end{equation*}
and if $a \in \mathbb{Z}[\theta]$, $a \neq 0$ and $a^{-1} = b/n$ with $b \in \mathbb{Z}[\theta]$ and $n \in \mathbb{Z}$, then:
\begin{align*}
	\s(b) &\leq (m-1) \s(a) + (m-1) \log C_1 + \log C_2 + \frac{1}{2} \log m, \\
	\log |n| &\leq m \s(a) + m \log C_1 - \frac{m}{2} \log m.
\end{align*}
\end{lemma}

Substituting our bounds on $C_1$, $C_2$ and $C_3$ from (\ref{C1}), (\ref{C2}) and (\ref{C3}) gives the following:

\begin{lemma} \label{lem:mult-size}
If $a, b \in \mathbb{Z}[\theta]$, then:
\begin{equation} \label{mul-size}
	\s(a b) \leq \s(a) + \s(b) + (m-1) F + 2 \log m,
\end{equation}
and if $a \in \mathbb{Z}[\theta]$, $a \neq 0$ and $a^{-1} = b/n$ with $b \in \mathbb{Z}[\theta]$ and $n \in \mathbb{Z}$, then:
\begin{align}
	\label{invnum-size}
		\s(b) &\leq (m-1) \s(a) + \frac{3}{2} m(m-1) F + \frac{3}{2} m \log m, \\
	\label{invden-size} \log |n|
		&\leq m \s(a) + m^2 F.
\end{align}
\end{lemma}

The next result is an upper and a lower bound for the absolute value of an element in terms of its coefficient size.

\begin{lemma} \label{lem:abs-by-size}
If $a \in \mathbb{Z}[\theta]$ and $a \neq 0$:
\begin{align}
	\label{abs-by-size}
	\log |a| &< \s(a) + m F, \\
	\label{invabs-by-size}
	\log |a^{-1}| &< (m-1) \s(a) + \frac{3}{2} m^2 (F+1).
\end{align}
\end{lemma}

\begin{proof}
(\ref{abs-by-size}) follows from (\ref{theta-geom-bound}):
\begin{equation*}
	\log |a| = \log \left| \sum_{j=0}^{m-1} a_j \theta^j \right|
	\leq \s(a) + \log \sum_{j=0}^{m-1} |\theta|^j
	< \s(a) + m F.
\end{equation*}

(\ref{invabs-by-size}) comes from (\ref{abs-by-size}) and (\ref{invnum-size}):
\begin{align*}
	\log |a^{-1}|
	&= \log \frac{|b|}{|n|}
	\leq \log |b|
	< \s(b) + m F \leq \\
	&\leq (m-1) \s(a) + m \frac{3m-1}{2} F + \frac{3}{2} m \log m,
\end{align*}
and after some simplifications, we get (\ref{invabs-by-size}).
\end{proof}

Sometimes we need to work with determinants involving algebraic integers, so we derive the following lemma from the previous results.

\begin{lemma} \label{lem:det-size}
Let $A \in \mathbb{Z}[\theta]^{n \times n}$ be a matrix with entries $a_{ij}$. Then:
\begin{equation*}
	\s(\det A) \leq n \max_{i,j} \s(a_{ij}) + (n-1)\left( (m-1) F + 2 \log m \right) + n \log n.
\end{equation*}
\end{lemma}

\begin{proof}
The full expansion of the determinant is:
\begin{equation} \label{det-exp}
	\det A = \sum_{\sigma \in S_n} (-1)^{N(\sigma)} \prod_{i=1}^{n} a_{i,\sigma(i)},
\end{equation}
where $S_n$ is the set of all $n!$ permutations of $\{1,2,\ldots,n\}$, and $N(\sigma)$ is the number of inversions in $\sigma$. Applying (\ref{addn-size}):
\begin{equation*}
	\s(\det A)
	\leq \max_{\sigma \in S_n} \s\left( \prod_{i=1}^{n} a_{i,\sigma(i)} \right) + \log n!,
\end{equation*}
and the proof finishes by applying (\ref{mul-size}) repeatedly and using that $n! \leq n^n$.
\end{proof}

\subsection{Running time of field operations} \label{sec:alg-time}

In this section we give bounds on the running time of several operations in algebraic number fields. There are already some results in this topic: \cite[Sect.~5]{belabas} and \cite[Sect.~3]{comp-hnf}. However, \cite{belabas} presents only multiplication and not in the form we can use (e.g.\ it does not differentiate between $F$ and the input size), and the results of \cite{comp-hnf} are not general enough (e.g.\ considering only fast multiplication). Nevertheless, we will use some ideas from these works. And we also present some operations that neither of them discusses.

As the field elements are represented by integers, these calculations rely on the running time of integer operations, especially multiplication and division. But there are several different algorithms for multiplying and dividing arbitrarily large integers, and each have different time complexity. For the sake of generality, we give our results in terms of the complexity of integer multiplication, using the following notation, similar to that in \cite{belabas}.

Let $\mul(A, B)$ be the running time of multiplying two integers $a, b \in \mathbb{Z}$ whose bit length are bounded by $A$ and $B$ (i.e.\ $\log |a| \leq A$ and $\log |b| \leq B$), and let $\mul(A) := \mul(A, A)$. The value depends on the actual integer multiplication algorithm used, for example:
\begin{itemize}
	\item basic multiplication: $\mul(A, B) = O(A B)$,
	\item Karatsuba multiplication: $\mul(A) = O(A^{\log_2 3})$,
	\item Schönhage--Strassen algorithm: $\mul(A) = O(A \log A \log \log A)$.
\end{itemize}
For more details about integer multiplication, see e.g.\ \cite[4.3.3.]{knuth-2}. In this paper we use only the following assumptions about the $\mul$ function:
\begin{align*}
	\mul(A, B) &= \mul(B, A), \\
	B \leq C &\Rightarrow \mul(A, B) \leq \mul(A, C), \\
	\mul(A, B + C) &\leq \mul(A, B) + \mul(A, C), \\
	\mul(A, nB) &\leq n \mul(A, B) \quad (n \in \mathbb{Z}^+), \\
	n \mul(A) &\leq \mul(n A), \\
	A \leq \mul(A) &\leq A^2.
\end{align*}

We assume furthermore that the integer division $a/b$ with quotient $q$ and remainder $r$ can be performed in $O(\mul(\log|b|, \log|q|))$ time (see e.g.\ \cite[4.3.3.~D]{knuth-2}).

We denote by $\T(expr)$ the number of arithmetic operations on machine words to calculate the expression $expr$. First we consider the most trivial operations:
\begin{lemma} \label{lem:triv-time}
If $a, b \in \mathbb{Z}[\theta]$ and $n \in \mathbb{Z}$, then
\begin{align}
	\label{add-time} \T(a \pm b) &= O \left( m \s(a, b)) \right), \\
	\label{intmul-time} \T(n a) &= O \left( m \mul(\s(n), \s(a)) \right).
\end{align}
\end{lemma}

For multiplication, we have the following result:

\begin{lemma} \label{lem:mul-time}
If $a, b \in \mathbb{Z}[\theta]$, then
\begin{equation} \label{mul-time}
	\T(a b) = O \left( m^2 \mul(\s(a), \s(b)) + m^2 \mul(m F,\, \s(a) + \s(b) + \log m ) \right).
\end{equation}
\end{lemma}

\begin{proof}
The product $c = a b$ can be computed by the following steps:
\begin{enumerate}
	\item Calculate the product as if $\theta$ were a symbol, i.e. perform a polynomial multiplication:
		\begin{equation*}
			d_l := \sum_j a_j b_{l-j} \quad (0 \leq l \leq 2m-2).
		\end{equation*}
	\item Calculate its remainder modulo $f$:
		\begin{equation*}
			c_l := d_l + \sum_{k=0}^{m-2} d_{m+k} r_{k,l} \quad (0 \leq l \leq m-1),
		\end{equation*}
		where $r_{k,l}$ are defined by (\ref{r-def}), and they can be precalculated from $f$ by (\ref{r-rec}).
\end{enumerate}
Both steps are dominated by the $O(m^2)$ multiplications, so the total running time is:
\begin{equation*}
	\T(a b) = O \left( m^2 \mul(\s(a), \s(b)) + m^2 \mul(\log \max_l |d_l|, \log \max_{k,l} |r_{k,l}|) \right).
\end{equation*}
To bound the variables in the second $\mul(\cdot)$, we use the definition of $d_l$ above and (\ref{r-bound}) for $r_{k,l}$:
\begin{align*}
	\log |d_l|
	&\leq \log \left( \sum_{j} |a_j| |b_{l-j}| \right)
	\leq \s(a) + \s(b) + \log m, \\
	\log |r_{k,l}| &\leq (k+1) F \leq m F,
\end{align*}
and substituting these gives the statement.
\end{proof}

There is another method, used by \cite{comp-hnf}, to calculate $c = a b$: we calculate a matrix $M_a$ from $a$ which turns $b$ into $c$, i.e.\ $M_a (b_0, b_1, \ldots, b_{m-1})^T = (c_0, c_1, \ldots, c_{m-1})^T$. In our case, it can be derived from the formulas above that:
\begin{equation} \label{ma-def}
	(M_a)_{i,j} = a_{i-j} + \sum_{k=0}^{j-2} a_{m+k-j+1} r_{k,i-1},
\end{equation}
(where $a_{i-j}$ is meant to be $0$ when $i < j$). Calculating $M_a$ therefore requires
\begin{equation} \label{ma-time}
	\T(M_a) = O\left( m^3 \mul(\s(a), m F) \right)
\end{equation}
time. This is already slower than our method with (\ref{mul-time}), but if we often multiply by the same $a$, we can calculate $M_a$ only once, and then the actual multiplication is just a matrix-vector multiplication, which has a slightly better complexity:
\begin{equation*}
	\T(a b)_{M_a} = O\left( m^2 \mul(\s(a) + m F, \s(b)) \right).
\end{equation*}

There is also another use of $M_a$, when calculating the inverse of $a$.

\begin{lemma} \label{lem:inv-time}
If $a \in \mathbb{Z}[\theta]$ and $a \neq 0$, then:
\begin{equation} \label{inv-time}
	\T(a^{-1}) = O \left( m^3 \mul(m \s(a) + m^2 F) \right).
\end{equation}
\end{lemma}

\begin{proof}
We follow the ideas of \cite[Prop.~10]{comp-hnf}, though using the more general $\mul$ function and the more precise $O(\cdot)$ notation. The calculation of $a^{-1} = b/n$ goes as follows:
\begin{enumerate}
	\item Calculate a matrix $M_a$ from $a$ as described above.
	\item Solve the linear system of equations $M_a (b_0/n, b_1/n, \ldots, b_{m-1}/n)^T = (1, 0, \ldots, 0)$ for $b/n$.
\end{enumerate}
The latter can be done e.g.\ by the Bareiss algorithm \cite{bareiss}, which runs in the following time, if $A$ is a bound on the matrix entries:
\begin{equation*}
	\T(\text{Bareiss}) = O\left( m^3 \mul(m \log(mA)) \right).
\end{equation*}

%

From the formula of $M_a$ (\ref{ma-def}), we can get that:
\begin{equation*}
	\log A = \log \max_{i,j} \left| (M_a)_{i,j} \right| = O(\s(a) + m F).
\end{equation*}
Substituting this already gives (\ref{inv-time}), and $\T(M_a)$ (\ref{ma-time}) has lower order.
\end{proof}

As a consequence, we can also calculate the running time of division:
\begin{lemma} \label{lem:div-time}
If $a, b \in \mathbb{Z}[\theta]$ and $b \neq 0$, then:
\begin{equation} \label{div-time}
	\T\left(\frac{a}{b}\right) = O \left( m^3 \mul(\s(a) + m \s(b) + m^2 F) \right).
\end{equation}
\end{lemma}

\begin{proof}
If $b^{-1} = \frac{c}{n}$ with $n \in \mathbb{Z}$, then $\frac{a}{b} = \frac{ac}{n}$. By (\ref{invnum-size}), $\s(c) = O\left( m \s(b) + m^2 F \right)$, and then $\T(b^{-1})$ by (\ref{inv-time}) and $\T(a c)$ by (\ref{mul-time}) gives the statement.
\end{proof}

In the rest of this section, we discuss some operations special to $\mathbb{R}$. Therefore, from now on we assume that $\theta \in \mathbb{R}$ and so $K \subset \mathbb{R}$.

First, we need the following auxiliary statement.
\begin{lemma} \label{lem:poly-approx}
Let $a, b \in \mathbb{Z}[\theta] \setminus \{0\}$. Let $\tilde\theta \in \mathbb{R}$ be so close to $\theta$ that
\begin{equation} \label{poly-approx-eps}
	\log |\tilde\theta - \theta|^{-1} \geq \s(a) + (m-1) \s(b) + \frac{3}{2} m(m+1) (F+1),
\end{equation}
and let $\tilde{a} = a_0 + a_1 \tilde\theta + a_2 \tilde\theta^2 + \ldots + a_{m-1} \tilde\theta^{m-1}$. Then:
\begin{equation*}
	|\tilde{a} - a| < |b|.
\end{equation*}
\end{lemma}

\begin{proof}
First we use only that $m |\tilde\theta - \theta| < \frac{1}{2}$ from (\ref{poly-approx-eps}).
\begin{align*}
	|\tilde{a} - a|
	&= \left| \sum_{k=0}^{m-1} a_k \left( \left(\theta + (\tilde\theta - \theta)\right)^k - \theta^k \right) \right|
	= \left| \sum_{k=0}^{m-1} a_k \sum_{j=1}^k \binom{k}{j} \theta^{k-j} (\tilde\theta - \theta)^j \right| \leq \\
	&\leq \sum_{k=0}^{m-1} |a_k| \sum_{j=1}^k \binom{k}{j} |\theta|^{k-j} |\tilde\theta - \theta|^j
	= \sum_{j=1}^{m-1} \sum_{l=0}^{m-j-1} \!|a_{l+j}|\, \binom{l+j}{j} |\theta|^l |\tilde\theta - \theta|^j \leq \\
	&\leq \left( \max_{i=0}^{m-1} |a_i| \right) \sum_{j=1}^{m-1} m^j |\tilde\theta - \theta|^j \sum_{l=0}^{m-j-1} |\theta|^l
	< \left( \max_{i=0}^{m-1} |a_i| \right) 2 m |\tilde\theta - \theta| \sum_{l=0}^{m-1} |\theta|^l.
\end{align*}
Taking logarithm and using (\ref{theta-geom-bound}), we get:
\begin{equation*}
	\log |\tilde{a} - a|
	< \s(a) + \log 2 m |\tilde\theta - \theta| + m F
	\leq \s(a) + \frac{3}{2} m (F+1) + \log |\tilde\theta - \theta|.
\end{equation*}
We add this inequality to (\ref{invabs-by-size}) for $b$:
\begin{equation*}
	\log |\tilde{a} - a| + \log |b^{-1}|
	< \s(a) + (m-1) \s(b) + \frac{3}{2} m(m+1) (F+1) + \log |\tilde\theta - \theta|.
\end{equation*}
By (\ref{poly-approx-eps}), the right-hand-side is $\leq 0$, so the left-hand-side is $< 0$, which is equivalent to the statement.
\end{proof}

Now we are able to calculate the worst-case complexity of the following operations.

\begin{lemma} \label{lem:cmp-time}
If $a, b \in \mathbb{Z}[\theta] \subset \mathbb{R}$, then:
\begin{equation} \label{cmp-time}
	\T(a < b) = \T(a \leq b) = O \left( m^2 \mul(m \s(a, b) + m^2 F) \right),
\end{equation}
\end{lemma}

\begin{proof}
Since $a < b$ is equivalent to $a - b < 0$, we need to consider only the $a < 0$ comparison, and the same is true for $\leq$. Assume the nontrivial case $a \neq 0$.

We approximate $\theta$ by a rational number $\tilde\theta = \frac{u}{d}$ where $d \in \mathbb{Z}^+$ and $u$ is either $\lfloor \theta d \rfloor$ or $\lceil \theta d \rceil$, the one with the smaller absolute value. We need so large $d$, hence so close $\tilde\theta$ to $\theta$, that the approximation $\tilde{a} = a_0 + a_1 \tilde\theta + a_2 \tilde\theta^2 + \ldots + a_{m-1} \tilde\theta^{m-1}$ and the exact $a$ have the same sign. This can be guaranteed if $|\tilde{a} - a| < |a|$, because then
\begin{align*}
	a > 0 &\implies \tilde{a} \geq a - |\tilde{a} - a| > 0, \\
	a < 0 &\implies \tilde{a} \leq a + |\tilde{a} - a| < 0.
\end{align*}
Now by Lemma~\ref{lem:poly-approx} with $b = a$, we can ensure this by choosing a $d$ with
\begin{equation*}
	\log d \geq m \s(a) + \frac{3}{2} m(m+1) (F+1),
\end{equation*}
because $\log |\tilde\theta - \theta|^{-1} > \log d$. We can choose $d$ to be the smallest such integer, but it is more efficient to round up to the nearest power of two. In any case,
\begin{equation} \label{den-size}
	\log d = O \left( m \s(a) + m^2 F \right).
\end{equation}

Now the approximation can be written as:
\begin{equation*}
	\tilde{a} = a_0 + a_1 \frac{u}{d} + \ldots + a_{m-1} \left(\frac{u}{d}\right)^{m-1} = \frac{ a_0 d^{m-1} + a_1 u d^{m-2} + \ldots + a_{m-1} u^{m-1} }{d^{m-1}}.
\end{equation*}
Call the numerator $r$. It can be calculated by the following recursion:
\begin{equation*}
	r_0 := 0, \qquad
	r_{k+1} := r_k u + a_{m-k-1} d^k, \qquad
	r := r_m.
\end{equation*}
Since $d$ is a power of two, the dominating operation is the multiplication $r_k u$.

We can bound $u$ and $r_k$ as follows, the latter by induction:
\begin{align*}
	|u| &= \min(|\lfloor \theta d \rfloor|, |\lceil \theta d \rceil|) \leq |\theta d| = |\theta| d, \\
	|r_k| &\leq \left( \max_{i=0}^{m-1} |a_i| \right) d^{k-1} \left( 1 + |\theta| + \ldots + |\theta|^{k-1} \right).
\end{align*}
Therefore, using (\ref{theta-geom-bound}) and (\ref{den-size}):
\begin{align*}
	\T(r_k u) &= \mul(\log |r_k|, \log |u|)
	\leq \mul(\s(a) + m \log d + m F, F + \log d) = \\
	&= O(\mul(m^2 \s(a) + m^3 F, m \s(a) + m^2 F))
	= O(m \mul(m \s(a) + m^2 F)),
\end{align*}
and the whole calculation performs $m$ such multiplications.
\end{proof}

\begin{lemma} \label{lem:round-time}
If $a, b \in \mathbb{Z}[\theta] \subset \mathbb{R}$, $n \in \mathbb{Z}$ and $b \neq 0$, $n \neq 0$, then:
\begin{align}
	\label{round-time}
	&\T\left( \left\lfloor \frac{a}{n} \right\rfloor \right) =
	\T\left( \left\lceil  \frac{a}{n} \right\rceil  \right) =
	\T\left( \left\lfloor \frac{a}{n} \right\rceil  \right) =
	O \left( m^2 \mul(m \s(a, n) + m^2 F) \right), \\
	\label{div-round-time}
	&\T\left( \left\lfloor \frac{a}{b} \right\rfloor \right) =
	\T\left( \left\lceil  \frac{a}{b} \right\rceil  \right) =
	\T\left( \left\lfloor \frac{a}{b} \right\rceil  \right) =
	O \left( m^2 \mul(m \s(a) + m^2 \s(b) + m^3 F) \right).
\end{align}
\end{lemma}

\begin{proof}
Since $\lfloor a/n \rceil = \lfloor a/n + 1/2 \rfloor$, we need to consider only $\lfloor \cdot \rfloor$ and $\lceil \cdot \rceil$. If $a \in \mathbb{Z}$, we can use integer division in $O(\mul(\s(a), \s(n)))$ time, so assume that $a \notin \mathbb{Z}$.

The proof of (\ref{round-time}) goes similarly to that of Lemma~\ref{lem:cmp-time}. Here the denominator $d$ must be so large that $\tilde{a}/n$ and $a/n$ have the same integer part. This can be ensured by
\begin{equation*}
	\left| \frac{\tilde{a}}{n} - \frac{a}{n} \right| <
	\min \left(
		\left| \frac{a}{n} - \left\lfloor \frac{a}{n} \right\rfloor \right|,
		\left| \frac{a}{n} - \left\lceil  \frac{a}{n} \right\rceil  \right|
	\right).
\end{equation*}
After multiplying this by $|n|$, we can use Lemma~\ref{lem:poly-approx} with
$b \in \big\{ a - n \lfloor a/n \rfloor,\, a - n \lceil a/n \rceil \big\}$,
but first we need to calculate $\s(b)$. Since $b - a$ is an integer, the coefficients of $a$ and $b$ are the same except the constant term, so $\s(b) \leq \max(\s(a), \log |b_0|)$. We can bound $|b_0|$ as follows (note that $|b/n| < 1$, so $|b| < |n|$):
\begin{equation*}
	|b_0| = |a_0 - a + b| < |a - a_0| + |n|
	= \left| \sum_{k=1}^{m-1} a_k \theta^k \right| + |n|
	\leq \max \left( |n|, \max_{i=1}^{m-1} |a_i| \right) \left( \sum_{i=0}^{m-1} |\theta|^i \right),
\end{equation*}
whose logarithm is bounded by $\s(a, n) + m F$, so:
\begin{equation*}
	\s(b) \leq \s(a, n) + m F.
\end{equation*}
Now by Lemma~\ref{lem:poly-approx}, we can choose $d$ such that
\begin{equation*}
	\log d = O \left( m \s(a, n) + m^2 F \right).
\end{equation*}
We can calculate in the same way as in Lemma~\ref{lem:cmp-time} that for $r$, the numerator of $\tilde{a}$:
\begin{align*}
	\T(r) &= O\left( m^2 \mul(m \s(a, n) + m^2 F) \right), \\
	|r| &= O\left( m^2 \s(a, n) + m^3 F \right).
\end{align*}
The last step is to divide it by $n$. This is an integer division, and takes $\mul(m^2 \s(a, n) + m^3 F, \s(n))$ time. Adding this to $\T(r)$ we get (\ref{round-time}).

For (\ref{div-round-time}), first we calculate $a/b$ in the form $ac/n$ with $n \in \mathbb{Z}$. This takes $\T(a/b) = O\left( m^3 \mul(\s(a) + m \s(b) + m^2 F) \right)$ time by Lemma~\ref{lem:div-time}. The length of the parameters are, by (\ref{invnum-size}), (\ref{mul-size}) and (\ref{invden-size}):
\begin{align*}
	\s(c) &= O \left( m \s(b) + m^2 F \right), \\
	\s(a c) &= O \left( \s(a) + \s(c) + m F \right)
		= O \left( \s(a) + m \s(b) + m^2 F \right), \\
	\log |n| &= O \left( m \s(b) + m^2 F \right).
\end{align*}
Putting these into (\ref{round-time}) and adding $\T(a/b)$ gives the result.
\end{proof}

\subsection{Summary of the operations} \label{sec:alg-summary}

The following table summarizes the results of Section~\ref{sec:alg} on the time complexity and the size of the results of the operations in $\mathbb{Z}[\theta]$. In this table, $a, b, c \in \mathbb{Z}[\theta]$ and $n \in \mathbb{Z}$.

\begin{center}
\begin{tabular}{|c|l|l|l|}
	\hline
	Operation & Output size & Time 
	\\ \hline
		$a \pm b$
		& $\s(a, b) + \log 2$
		& $O\left( m \s(a, b) \right)$
	\\ \hline
		$n a$
		& $\s(a) + \s(n)$
		& $O\left( m \mul(\s(n), \s(a)) \right)$
	\\ \hline
		$a b$
		& $O\left( \s(a) + \s(b) + m F \right)$
		& \begin{tabular}{@{}l@{}} $O\,\big( m^2 \mul(\s(a), \s(b))\; + $ \\ $ \phantom{O\,\big(} m^2 \mul(m F,\, \s(a) + \s(b) + \log m ) \big)$ \end{tabular}
	\\ \hline
		$a^{-1} \rightarrow \frac{b}{n}$
		& $O\left( m \s(a) + m^2 F \right)$
		& $O\left( m^3 \mul(m \s(a) + m^2 F) \right)$
	\\ \hline
		$\frac{a}{b} \rightarrow \frac{c}{n}$
		& $O\left( \s(a) + m \s(b) + m^2 F \right)$
		& $O\left( m^3 \mul(\s(a) + m \s(b) + m^2 F) \right)$
	\\ \hline
		$a < b$, $a \leq b$
		& $O\left( 1 \right)$
		& $O\left( m^2 \mul(m \s(a, b) + m^2 F) \right)$
	\\ \hline
		$\left\lfloor \frac{a}{n} \right\rfloor,
		\left\lceil \frac{a}{n} \right\rceil,
		\left\lfloor \frac{a}{n} \right\rceil$
		& $O\left( \s(a) + m F \right)$
		& $O\left( m^2 \mul(m \s(a, n) + m^2 F) \right)$
	\\ \hline
		$\left\lfloor \frac{a}{b} \right\rfloor,
		\left\lceil \frac{a}{b} \right\rceil,
		\left\lfloor \frac{a}{b} \right\rceil$
		& $O\left( s(a) + m s(b) + m^2 F \right)$
		& $O\left( m^2 \mul(m \s(a) + m^2 \s(b) + m^3 F) \right)$
	\\ \hline
\end{tabular}
\end{center}

\section{Bareiss algorithm} \label{sec:bareiss}

The Bareiss algorithm \cite{bareiss} is an integer-preserving modification of Gaussian elimination for $\mathbb{Z}$ that maintains as small integers as generally possible by using provably exact divisions to reduce their sizes. In this section we apply the algorithm to $\mathbb{Z}[\theta]$ (where $\theta$ is not neccessarily real) and calculate its running time using the results of the previous section, and compare it with the running time in $\mathbb{Z}$. We consider the simpliest form of the algorithm, when a square matrix is converted into an upper triangular form (e.g.\ to calculate its determinant).

Let $M \in \mathbb{Z}[\theta]^{n \times n}$ with elements $a_{ij}$, and let $A := \max_{i,j} \s(a_{ij})$. The Bareiss algorithm uses the following formula \cite[p.~570]{bareiss}:
\begin{equation} \label{bareiss-rec}
	a_{00}^{(0)} = 1, \quad
	a_{ij}^{(1)} = a_{ij}, \qquad
	a_{ij}^{(k+1)} = \frac{a_{kk}^{(k)} a_{ij}^{(k)} - a_{ik}^{(k)} a_{kj}^{(k)}}{a_{k-1,k-1}^{(k-1)}}
\end{equation}
with $1 \leq k \leq n-1$ and $k+1 \leq i,j \leq n$.

First we bound the size of all intermediate variables, $a_{ij}^{(k)}$. They can be written as determinants of order $k$ ($\leq n$) whose entries are from $M$ \cite[p.~565]{bareiss}. Therefore, the division in (\ref{bareiss-rec}) is exact and $a_{ij}^{(k)} \in \mathbb{Z}[\theta]$, so we can use $\s(\cdot)$ to measure their size. By Lemma~\ref{lem:det-size}, we have:
\begin{equation} \label{bareiss-size}
	B := \max_{i,j,k} \s\left(a_{ij}^{(k)}\right) = O(n A + n m F + n \log n).
\end{equation}

Now we estimate the running time of the recursive formula (\ref{bareiss-rec}). The calculation consists of the following main operations:
\begin{enumerate}
	\item two multiplications:
	\begin{itemize}
		\item time: $O\left( m^2 \mul(B) \right)$ by (\ref{mul-time}),
		\item output size: $O(B)$ by (\ref{mul-size});
	\end{itemize}
	\item exact division:
	\begin{enumerate}
		\item calculating $\left( a_{k-1,k-1}^{(k-1)} \right)^{-1}$ in the form $b_{k-1,k-1}^{(k-1)} \big/ d_{k-1,k-1}^{(k-1)}$:
		\begin{itemize}
			\item time: $O\left( m^3 \mul(m B) \right)$ by (\ref{inv-time}),
			\item output size: $O(m B)$ by (\ref{invnum-size}) and (\ref{invden-size});
		\end{itemize}
		\item multiplying the numerator by $b_{k-1,k-1}^{(k-1)}$:
		\begin{itemize}
			\item time: $O\left( m^2 \mul(B, m B) \right) = O\left( m^3 \mul(B) \right)$ by (\ref{mul-time}),
			\item output size: $O(m B)$ by (\ref{mul-size});
		\end{itemize}
		\item and dividing the resulting algebraic number exactly by the integer $d_{k-1,k-1}^{(k-1)}$:
		\begin{itemize}
			\item time: $O\left( m \mul(m B) \right)$.
		\end{itemize}
	\end{enumerate}
\end{enumerate}

Most of these are done for each $i,j,k$, i.e.\ $O(n^3)$ times, but 2/(a) depends only on $k$, so it is done $O(n)$ times. Adding all these together, we have:
\begin{equation*}
	\T_{\mathbb{Z}[\theta]}(\text{Bareiss})
	= O\left( n^3 m^3 \mul(B) + n m^3 \mul(m B) \right).
\end{equation*}
For comparison, the running time over $\mathbb{Z}$ is:
\begin{equation*}
	\T_{\mathbb{Z}}(\text{Bareiss})
	= O\left( n^3 \mul(n A + n \log n)) \right).
\end{equation*}

We can see that if we ignore the field-dependent constants $m$ and $F$, the running time is asymptotically the same. In the following table, we compare several different $\mul(\cdot)$ functions (see Section~\ref{sec:alg-time}) for both cases. In the last row, we introduced a simplified notation $\tilde{O}(\ldots)$ for omitting the logarithmic factors, i.e.\ $\tilde{O}(N) = O(N \log N \log \log N)$.

\begin{center}
\begin{tabular}{|l|l|l|}
	\hline
	$\mul(X)$
	& $\T_{\mathbb{Z}[\theta]}(\text{Bareiss})$
	& $\T_\mathbb{Z}(\text{Bareiss})$
	\\ \hline
	$X^2$
	& $O\left( n^3 m^3 (n^2 + m^2) (A + m F + \log n)^2 \right)$
	& $O\left( n^5 (A + \log n)^2 \right)$
	\\ \hline
	$X^{\log_2 3}$
	& $O\left( n^{2.6} m^3 (n^2 + m^{1.6}) (A + m F + \log n)^{1.6} \right)$
	& $O\left( n^{4.6} (A + \log n)^{1.6}) \right)$
	\\ \hline
	$X \log X \log \log X$
	& $\tilde{O}\left( n^2 m^3 (n^2 + m) (A + m F + \log n) \right)$
	& $\tilde{O}\left( n^4 (A + \log n) \right)$
	\\ \hline
\end{tabular}
\end{center}

\section{LLL algorithm} \label{sec:lll}

The LLL algorithm is a lattice basis reduction algorithm invented by Lenstra, Lenstra and Lovász \cite{lll}. It is known that it runs in polynomial time if the vectors are in $\mathbb{Z}^n$. In this section we show that it is also polynomial for vectors in $\mathbb{Z}[\theta]^n$ for real $\theta$.

Let $b_1, b_2, \ldots, b_n \in \mathbb{R}^n$ be a basis. Then
\begin{equation*}
	\Lambda(b_1, b_2, \ldots, b_n) := \Big\{ c_1 b_1 + c_2 b_2 + \ldots + c_n b_n \;\big|\; c_1, c_2, \ldots, c_n \in \mathbb{Z} \Big\}
\end{equation*}
is called the \emph{lattice} spanned by $b_1, b_2, \ldots, b_n$.

The LLL algorithm modifies $b_1, b_2, \ldots, b_n$ step by step, preserving the spanned lattice, and finally turning the vectors to a reduced basis (in the sense defined below). At any point in the algorithm, we define the Gram--Schmidt orthogonalization of the actual $b_i$ vectors as follows:
\begin{align}
	\label{def:ob} b_i^* &:= b_i - \sum_{j=1}^{i-1} \mu_{ij}b_j^* & &(1 \leq i \leq n), \\
	\label{def:mu} \mu_{ij} &:= \frac{\langle b_i, b_j^*\rangle}{\langle b_j^*, b_j^*\rangle} & &(1 \leq j < i \leq n).
\end{align}
When the algorithm terminates, the $b_i$ vectors are LLL-reduced, which means the following two properties:
\begin{align}
	\label{mu-cond} \left| \mu_{ij} \right| &\leq \frac{1}{2} & &(1 \leq j < i \leq n), \\
	\label{size-cond} \left\lVert b_i^* + \mu_{i\,i-1} b_{i-1}^* \right\rVert_2^2 &\geq \delta \left\lVert b_{i-1}^* \right\rVert_2^2 & &(2 \leq i \leq n),
\end{align}
where $\delta$ is a parameter of the algorithm between $\frac{1}{4} < \delta < 1$, often $\delta = \frac{3}{4}$.

The skeleton of the LLL algorithm is the following. This contains only the changes of $b_i$'s. The full algorithm keeps track of other variables after each $b_i$-change to maintain (\ref{def:ob}) and (\ref{def:mu}).
\begin{algorithm}
$k := 2$\;
\While{$k \leq n$}{
	$b_k := b_k - \left\lfloor \mu_{k\,k-1} \right\rceil b_{k-1}$\;
	\eIf{$k \geq 2 \wedge \left\lVert b_k^* + \mu_{k\,k-1} b_{k-1}^* \right\rVert_2^2 < \delta \left\lVert b_{k-1}^* \right\rVert_2^2$}{
		\emph{(swap step)}\;
		$b_k \leftrightarrow b_{k-1}$\;
		$k := k - 1$\;
	}{
		\emph{(reduction step)}\;
		\For{$l := k-2$ \KwTo $1$}{
			$b_k := b_k - \left\lfloor \mu_{kl} \right\rceil b_l$\;
		}
		$k := k + 1$\;
	}
}
\end{algorithm}

\subsection{Properties in $\mathbb{R}^n$} \label{sec:lll-basic}

First we discuss some properties of the LLL algorithm on any real basis $b_1, b_2, \ldots, b_n \in \mathbb{R}^n$. We start with some definitions.

Given the lattice $\Lambda := \Lambda(b_1, b_2, \ldots, b_n)$, we need the shortest vector length in $\Lambda$:
\begin{equation*}
	L_0 := \min \left\{ \lVert x \rVert_2^2 \;\big|\; x \in \Lambda \setminus \{0\} \right\}.
\end{equation*}

Other important quantities are the numbers $d_1, d_2, \ldots, d_n \in \mathbb{R}$, which depend on the $b_i$ vectors, and can be defined by any of the following equivalent expressions \cite[p.~521]{lll}:
\begin{align}
	\label{def:dl-obi} d_l &= \lVert b_1^* \rVert_2^2 \, \lVert b_2^* \rVert_2^2 \ldots \lVert b_l^* \rVert_2^2, \\
	\label{def:dl-det} d_l &= \det\left( \langle b_i, b_j \rangle \right)_{1 \leq i,j \leq l}, \\
	\label{def:dl-lambda} d_l &= \det(\Lambda(b_1, b_2, \ldots, b_l))^2.
\end{align}

For convenience, we extend this to $d_0 := 1$.

We will need the following inequalities between these quantities, which are independent from the context of the LLL algorithm.

\begin{lemma} \label{lem:latice-ineq}
\begin{align}
	\label{dl-lower} & & d_l &\geq \left( \frac{L_0}{l} \right)^l, & (1 \leq l \leq n)& \\
	\label{L0-bounds} & & \min_{i=1}^n \,\lVert b_i^* \rVert_2^2 &\leq L_0 \leq \lVert b_1^* \rVert_2^2.
\end{align}
\end{lemma}

\begin{proof}
Minkowski's theorem \cite[III.2.2.]{geom-num} states that if $S \subseteq \mathbb{R}^l$ is convex, symmetric to the origin, and has no other common point with the $\Lambda_l := \Lambda(b_1, b_2, \ldots, b_l)$ lattice than the origin, then:
\begin{equation*}
	\det(\Lambda_l) \geq 2^{-l} \operatorname{Vol}(S),
\end{equation*}
where $\operatorname{Vol}(S)$ volume of $S$. Applying this to a hypercube with side $2r/\!\sqrt{l}$ with $r < \sqrt{L_0}$ (note that its circumscribed sphere has radius $r$), then by $r \rightarrow \sqrt{L_0}$ and squaring, we get (\ref{dl-lower}).

The upper bound on $L_0$ in (\ref{L0-bounds}) is a consequence of this, since $d_1 = \lVert b_1^* \rVert_2^2$. The lower bound on $L_0$ is elementary, and follows e.g.\ from the proof of \cite[(1.11)]{lll}.
\end{proof}

The following lemma shows that after any whole number of iterations in the LLL algorithm, the variables and related quantities can be bounded by initially known expressions.

\begin{lemma} \label{lem:lll-bounds}
Let $B := \max_{i=1}^n \,\lVert b_i \rVert_2^2$ for the input vectors in the LLL algorithm. Then, at the beginning or end of the body of the main {\bf while}-loop, the following inequalities hold, depending on the loop variable $k$:
\begin{align}
	\label{obi-bound} \lVert b_i^* \rVert_2^2 &\leq B, & \\
	\label{bi-bound} \lVert b_i \rVert_2^2 &\leq nB & &(i \neq k), \\
	\label{mu-half} |\mu_{ij}| &\leq \frac{1}{2} & &(i < k), \\
	\label{mukj-bound} |\mu_{ij}| &\leq 2^{n-i} \!\sqrt{n} \left( \frac{nB}{L_0} \right)^{\frac{n-1}{2}} & &(i = k), \\
	\label{muij-bound} |\mu_{ij}| &\leq \sqrt{n} \left( \frac{jB}{L_0} \right)^{\frac{j}{2}} & &(i > k), \\
	\label{dl-bound} d_j &\leq B^j. &
\end{align}
\end{lemma}

\begin{proof}
These inequalities are analogous to \cite[p.~523]{lll}, with the difference that \cite{lll} works in $\mathbb{Z}^n$, so it can use the fact that $d_l \geq 1$ since $d_l$ is both integer and positive. We replace this by the more general (\ref{dl-lower}).

(\ref{obi-bound}), (\ref{bi-bound}), (\ref{mu-half}) and (\ref{dl-bound}) are the same as in \cite{lll}.

We prove (\ref{muij-bound}) by using our other inequalities and the Cauchy--Schwarz inequality:
\begin{equation*}
	|\mu_{ij}|^2
	\overset{(\ref{def:mu})}{=}
	\frac{\left| \langle b_i, b_j^* \rangle \right|^2}{\lVert b_j^* \rVert_2^4}
	\overset{\text{C.-S.}}{\leq}
	\frac{\lVert b_i \rVert_2^2 \lVert b_j^* \rVert_2^2}{\lVert b_j^* \rVert_2^4}
	\overset{(\ref{def:dl-obi})}{=}
	\frac{d_{j-1}}{d_j} \lVert b_i \rVert_2^2
	\overset{\substack{(\ref{dl-bound})\\(\ref{dl-lower})}}{\leq}
	\frac{B^{j-1}}{\left( \frac{L_0}{j} \right)^j} \lVert b_i \rVert_2^2
	\overset{(\ref{bi-bound})}{\leq}
	n \left( \frac{j B}{L_0} \right)^j.
\end{equation*}
Then the derivation of (\ref{mukj-bound}) from (\ref{muij-bound}) works in the same way as the analogue \cite[(1.34)]{lll}.
\end{proof}

Now we can give an estimate on the number of iterations in the LLL algorithm.

\begin{lemma} \label{lem:iter-bound}
Let $N$ be the number of main iterations (both reduction steps and swap steps), and let $K_\delta := \frac{1}{\log \frac{1}{\delta}}$. Then:
\begin{equation*}
	N = O\left( n^2 \log\frac{nB}{L_0} K_\delta \right).
\end{equation*}
\end{lemma}

\begin{proof}
Let $N_r$ be the number of reduction steps, and $N_s$ be the number of swap steps. Since a reduction adds, and a swap subtracts $1$ from $k$, and since the algorithm starts with $k = 2$ and finishes when $k = n+1$, therefore $N_r - N_s = n-1$, so $N = N_r + N_s = 2N_s + n-1$, i.e.\ it is sufficient to estimate $N_s$.

Let $D := d_1 d_2 \ldots d_n$, and let $D^{(s)}$ be the value of $D$ after $s$ swap steps.

\cite{lll} proves that for integer values (i.e.\ for $b_1, \ldots, b_n \in \mathbb{Z}^n$) there are at most $O(n^2 \log B)$ iterations (or rather $O(n^2 \log B\, K_\delta)$ if we want to capture $\delta$), but this uses the fact that $D$ is an integer, hence $D \geq 1$. We replace this with another lower bound for $D$. We also need an upper bound for $D$:
\begin{align*}
	D &= \prod_{j=1}^n d_j
	\overset{(\ref{dl-lower})}{\geq} \prod_{j=1}^n \left( \frac{L_0}{j} \right)^j
	\geq \prod_{j=1}^n \left( \frac{L_0}{n} \right)^j
	= \left( \frac{L_0}{n} \right)^{\frac{n(n+1)}{2}}, \\
	D &= \prod_{j=1}^n d_j
	\overset{(\ref{dl-bound})}{\leq} \prod_{j=1}^n B^j
	= B^{\frac{n(n+1)}{2}}.
\end{align*}
These bounds are true after any number of iterations, i.e.\ for any $D^{(s)}$. Furthermore, we use the fact that a reduction step does not change $D$, and that a swap step reduces $D$ by a factor $<\delta$: $D^{(s+1)} < \delta D^{(s)}$ -- both are proved in \cite[p.~521]{lll} without the use of the integer property. By induction, it follows that $D^{(s)} < \delta^s D^{(0)}$.
Putting these inequalities together:
\begin{equation*}
	\left( \frac{L_0}{n} \right)^{\frac{n(n+1)}{2}} \leq D^{(N_s)} < \delta^{N_s} D^{(0)} \leq \delta^{N_s} B^{\frac{n(n+1)}{2}}.
\end{equation*}
After taking logarithms from both ends and rearranging, we get:
\begin{equation*}
	N_s < \frac{1}{\log \frac{1}{\delta}} \frac{n(n+1)}{2} \log \frac{nB}{L_0},
\end{equation*}
and the statement follows from this, because $N = 2N_s + n-1$.
\end{proof}

\subsection{Coefficient size in $\mathbb{Z}[\theta]^n$} \label{lll-size}

Now we restrict the basis to be over a real number field, i.e.\ from now on, $b_1, b_2, \ldots, b_n \in \mathbb{Z}[\theta]^n \subset \mathbb{R}^n$. We use the notations $m$, $F$ and $\s(\cdot)$ as in Section~\ref{sec:alg}, and extend $\s(\cdot)$ naturally to vectors: $\s(x) := \max_{j=1}^n \s(x_j)$.

When implementing the LLL algorithm exactly, we do not need to maintain the non-integral quantities $b_i^*$ and $\mu_{ij}$. Instead, as presented e.g.\ in \cite[Alg.~2.6.7]{cohen} for $\mathbb{Z}$, we can use the integer $d_j$, and write $\mu_{ij} = \lambda_{ij} / d_j$ where $\lambda_{ij}$ is also an integer (see e.g.\ \cite[Prop.~2.6.5]{cohen}). The same applies to $\mathbb{Z}[\theta]$, i.e.\ $d_j, \lambda_{ij} \in \mathbb{Z}[\theta]$.

We need to give bounds on $\s(d_j)$ and $\s(\lambda_{ij})$ during the algorithm. First we do this in terms of the current $\s(b_i)$, outside of the context of the LLL algorithm.

\begin{lemma} \label{lem:d-size}
For the corresponding values of $b_i$, $d_j$ and $\lambda_{ij}$, we have:
\begin{equation*}
	\s\left(d_j, \lambda_{ij}\right)
	\leq 2n \left( \max_{i=1}^{n} \s(b_i) + (m-1)F + 2 \log m + \log n \right).
\end{equation*}
\end{lemma}

\begin{proof}
We already know from (\ref{def:dl-det}) that $d_j$ is a $j \times j$ determinant of elements like $\langle b_{i'}, b_{j'} \rangle$. We show that $\lambda_{ij}$ has a similar structure. This follows e.g. from the proof of \cite[Prop.~2.6.5]{cohen} showing that $\lambda_{ij}$ is integer, where we have
\begin{equation*}
\begin{pmatrix}
	\langle b_1, b_1 \rangle & \cdots & \langle b_1, b_j \rangle \\
	\vdots & \ddots & \vdots \\
	\langle b_j, b_1 \rangle & \cdots & \langle b_j, b_j \rangle
\end{pmatrix}
\begin{pmatrix}
	\xi_1 \\ \vdots \\ \xi_j
\end{pmatrix}
=
\begin{pmatrix}
	\langle b_1, b_i \rangle \\
	\vdots \\
	\langle b_j, b_i \rangle
\end{pmatrix}
\end{equation*}
with $\xi_j = \mu_{ij}$. Solving the system for $\mu_{ij} = \lambda_{ij} / d_j$ by Cramer's rule gives the needed determinant structure for both $\lambda_{ij}$ and $d_j$.

First we give an estimate on the coefficient size of the individual $\langle b_{i'}, b_{j'} \rangle$ elements using the properties of the $\s(\cdot)$ operator:
\begin{equation*}
	\s\left( \langle b_{i'}, b_{j'} \rangle \right)
	\leq \s(b_{i'}) + \s(b_{j'}) + (m-1)F + 2 \log m + \log n.
\end{equation*}
Then applying Lemma~\ref{lem:det-size} to the $j \times j$ determinants (using that $j \leq n$) finishes the proof.
\end{proof}

In Section~\ref{sec:lll-basic} about the LLL algorithm over $\mathbb{R}$, several inequalities used the constant $L_0$, mostly in the form $\frac{n B}{L_0}$. In $\mathbb{Z}[\theta]$, we can get rid of $L_0$, and use only basic parameters, including the coefficient size of the input vectors:
\begin{equation} \label{def:A}
	A := \max_{i=1}^{n} \s(b_i).
\end{equation}

\begin{lemma} \label{lem:H-bound}
Consider the LLL algorithm over $\mathbb{Z}[\theta]$, and let
\begin{equation} \label{def:H}
	H := \frac{1}{n} \log \frac{nB}{L_0},
\end{equation}
then we have:
\begin{equation} \label{H-bound}
	H = O\left( m A + m^2 F + m \log n \right).
\end{equation}
\end{lemma}

\begin{proof}
We can give the following lower bound on $L_0$ in terms of the initial $d_l$'s:
\begin{equation*}
	\frac{1}{L_0}
	\overset{(\ref{L0-bounds})}{\leq}
	\max_{i=1}^{n} \frac{1}{\lVert b_i^* \rVert_2^2}
	\overset{(\ref{def:dl-obi})}{=}
	\max_{i=1}^{n} \frac{d_{i-1}}{d_i},
\end{equation*}
and use Lemma~\ref{lem:abs-by-size} to bound $d_l$ from two sides by $\s(d_l)$, and use Lemma~\ref{lem:d-size} about $\s(d_l)$:
\begin{equation*}
	\log \frac{d_{i-1}}{d_i}
	= \log d_{i-1} + \log d_i^{-1}
	< m \max_{j=0}^{n} \s(d_j) + O(m^2 F)
	= O\left( n m A + n m^2 F + n m \log n \right).
\end{equation*}
It remains to give an upper bound on $B$ in terms of $A$, again by Lemma~\ref{lem:abs-by-size}:
\begin{equation*}
	\log B
	= \log \max_{i=1}^{n} \,\lVert b_i \rVert_2^2
	\leq 2 \log \max_{i=1}^{n} \,\lVert b_i \rVert_\infty + \log n
	< 2 A + 2 m F + \log n,
\end{equation*}
and we get the statement by combining these results.
\end{proof}

\begin{lemma} \label{lem:sbk-change}
Consider one main step of the LLL algorithm, either a reduction step or a swap step. If the vectors before and after the step are called $b_i$ and $b_i'$ respectively, then:
\begin{equation*}
	\s(b_i') \leq \max_{l=1}^{n} \s(b_l) + \frac{n^2}{2} H + n \log 2.
\end{equation*}
\end{lemma}

\begin{proof}
The following pseudocode shows the changes made to $b_i$ and $\mu_{ij}$ in a single reduction step. (Note that compared to the full algorithm presented earlier, here the $\mu_{kj}$'s are also present, and all changes of $b_k$ are joined into the $l$-loop. Also note that this is presented for the sake of the calculation, but we use the integral version of the algorithm instead, as described in Section~\ref{sec:lll-time}, which manipulates $\lambda_{ij}$ and $d_j$ instead of the rational $\mu_{ij}$.)

\begin{algorithm}[H]
\For{$l := k-1$ \KwTo $1$}{
	$b_k := b_k - \lfloor \mu_{kl} \rceil b_l$\;
	\For{$j := 1$ \KwTo $l-1$}{
		$\mu_{kj} := \mu_{kj} - \lfloor \mu_{kl} \rceil \mu_{lj}$\;
	}
	$\mu_{kl} := \mu_{kl} - \lfloor \mu_{kl} \rceil$\;
}
\end{algorithm}

A swap step makes fewer of these changes (only for $l = k-1$), and additionally an exchange of two $b_i$'s, which does not change the maximum of $\s(b_i)$. Therefore, we can concentrate on the reduction step only.

None of the $b_i$ or $\mu_{ij}$ change for $i \neq k$, so the statement trivially holds for these $b_i$. In order to calculate the change of $\s(b_k)$, we need the size of $\lfloor \mu_{kl} \rceil$ in the algorithm. To distinguish between different values of the variable, we call $\mu_{kl}$ the initial value, and $\mu_{kl}'$ the value when taking $\lfloor \cdot \rceil$. Examining the code above, we can see that
\begin{equation*}
	\mu_{kl}' = \mu_{kl} - \!\sum_{i=l+1}^{k-1} \lfloor \mu_{ki}' \rceil \mu_{il},
\end{equation*}
so, using the bounds on $\mu_{ij}$, (\ref{mu-half}) and (\ref{mukj-bound}):
\begin{equation*}
	\left|\lfloor \mu_{kl}' \rceil\right| \leq 2 |\mu_{kl}'|
	\leq 2 |\mu_{kl}| + 2 \!\sum_{i=l+1}^{k-1} \left|\lfloor \mu_{ki}' \rceil\right| |\mu_{il}|
	\leq 2^{n-k+1} \sqrt{n} \left( \frac{nB}{L_0} \right)^{\frac{n-1}{2}} + \sum_{i=l+1}^{k-1} \left|\lfloor \mu_{ki}' \rceil\right|.
\end{equation*}
From this, we can show by induction from $l = k-1$ to $1$ that
\begin{equation*}
	\left|\lfloor \mu_{kl}' \rceil\right| \leq 2^{n-l} \sqrt{n} \left( \frac{nB}{L_0} \right)^{\frac{n-1}{2}}\!.
\end{equation*}

Now we can calculate the change of $\s(b_k)$:
\begin{equation*}
	\s(b_k')
	= \s\left( b_k - \sum_{l=1}^{k-1} \lfloor \mu_{kl}' \rceil b_l \right)
	\leq \max_{l=1}^{k} \s(b_l) + \log \left( 1 + \sum_{l=1}^{k-1} \big| \lfloor \mu_{kl}' \rceil \big| \right).
\end{equation*}
We give an upper bound on the argument of this logarithm using our bound on $\lfloor \mu_{kl}' \rceil$:
\begin{align*}
	1 + \sum_{l=1}^{k-1} \big| \lfloor \mu_{kl}' \rceil \big|
	&\leq 1 + \sum_{l=1}^{k-1} 2^{n-l} \sqrt{n} \left( \frac{nB}{L_0} \right)^{\frac{n-1}{2}} = \\
	&= 1 + \left( 2^n - 2^{n-k+1} \right) n^{\frac{1}{2}} \left( \frac{nB}{L_0} \right)^{\frac{n-1}{2}}
	\leq 2^n \left( \frac{nB}{L_0} \right)^{\frac{n}{2}}\!.
\end{align*}

In the last step, we used that $\frac{nB}{L_0} \geq n$, which follows from $L_0 \leq B$ by (\ref{L0-bounds}) and (\ref{obi-bound}). Since the logarithm of the right-hand-side is $\frac{n^2}{2} H + n \log 2$ (see the definition of $H$: (\ref{def:H})), the proof is completed.
\end{proof}

Now we can combine all these results together to get the $\s(\cdot)$ of the main variables during the LLL algorithm.

\begin{lemma} \label{lem:lll-size}
In the LLL algorithm, at the beginning or end of the body of the main {\bf while}-loop, the following inequalities hold:
\begin{align}
	\label{bi-size} \s(b_i) &= O\left( n^5 H^2 K_\delta \right)\!, \\
	\label{dl-size} \s(d_j) &= O\left( n^6 H^2 K_\delta \right)\!, \\
	\label{lambda-size} \s(\lambda_{ij}) &= O\left( n^6 H^2 K_\delta \right)\!.
\end{align}
\end{lemma}

\begin{proof}
If we repeatedly apply Lemma~\ref{lem:sbk-change} for the first $t$ steps of the LLL algorithm, then:
\begin{equation*}
	\s(b_i) \leq A + O\left( n^2 H \right) t,
\end{equation*}
and because $t = O\left( n^3 H K_\delta \right)$ by Lemma~\ref{lem:iter-bound} and $H$ dominates $A$ by Lemma~\ref{lem:H-bound}, we get (\ref{bi-size}). The other two follows from this by Lemma~\ref{lem:d-size}.
\end{proof}

\subsection{Running time of the LLL algorithm} \label{sec:lll-time}

Now we have enough information to calculate an upper bound for the running time of the LLL algorithm for algebraic numbers. The basic structure of the algorithm is presented at the beginning of Section~\ref{sec:lll}, and for the details, we use \cite[Alg.~2.6.7]{cohen}, but adapted to $\mathbb{Z}[\theta]$ instead of $\mathbb{Z}$. The algorithm has three parts where significant operations take place:
\begin{enumerate}
	\item reduction with a single $\mu_{kl}$, i.e. the assignment $b_k := b_k - \lfloor \mu_{kl} \rceil b_l$ and related neccessary changes (note that this is not a complete reduction step, which does this $k-1$ times);
	\item exchange of $b_k$ and $b_{k-1}$ and the related neccessary changes;
	\item the comparison $\left\lVert b_k^* + \mu_{k\,k-1} b_{k-1}^* \right\rVert_2^2 < \delta \left\lVert b_{k-1}^* \right\rVert_2^2$.
\end{enumerate}
We denote the running time of these steps as $\T(\text{red})$, $\T(\text{swap})$ and $\T(\text{cmp})$ respectively.

Let $D$ be the bound on all $\s(d_j)$, $\s(\lambda_{ij})$ and $\s(b_i)$ after any number of iterations, and we know by Lemma~\ref{lem:lll-size} that $D = O\left( n^6 H^2 K_\delta \right)$.

First consider $\T(\text{red})$. Its crutial part is to calculate $q := \lfloor \mu_{kl} \rceil = \lfloor \lambda_{kl} / d_l \rceil$. This is an integer, and no matter how big $\s(\lambda_{kl})$ and $\s(d_l)$ were, it can be much smaller:
\begin{equation*}
	\log |q|
	= \log \big| \lfloor \mu_{kl} \rceil \big|
	\leq \log 2 |\mu_{kl}|
	\overset{(\ref{mukj-bound})}{\leq}
	\frac{n-1}{2} \log \frac{nB}{L_0} + \frac{1}{2} \log n + n \log 2
	\overset{(\ref{def:H})}{=}
	O\left( n^2 H \right).
\end{equation*}
On the left, we show all steps of the reduction, and on the right, we gave the complexities of the major operations (note that $D$ dominates $m F$, and that $b_l$ is a vector with $n$ components):
\begin{center}
\begin{tabular}{l l}
\parbox{0.35\linewidth}{
	\begin{algorithm}[H]
	$q := \left\lfloor \frac{\lambda_{kl}}{d_l} \right\rceil$\;
	$b_k := b_k - q b_l$\;
	\For{$j := 1$ \KwTo $l-1$}{
		$\lambda_{kj} := \lambda_{kj} - q \lambda_{lj}$\;
	}
	$\lambda_{kl} := \lambda_{kl} - q d_l$\;
	\end{algorithm}
} &
\parbox{0.50\linewidth}{
	$\T\left( \left\lfloor \frac{\lambda_{kl}}{d_l} \right\rceil \right)
	= O\left( m^2 \mul(m^2 D) \right)$
	by (\ref{div-round-time})
	
	$\T(q b_l)
	= O\left( n m \mul(n^2 H, D) \right)$
	by (\ref{intmul-time})
	
	$\T(q \lambda_{lj})
	= O\left( m \mul(n^2 H, D) \right)$
	by (\ref{intmul-time})
	
	$\T(q d_l)
	= O\left( m \mul(n^2 H, D) \right)$
	by (\ref{intmul-time})
}
\end{tabular}
\end{center}

The complexity of all these steps is:
\begin{align*}
	\T(\text{red})
	&= \T\left( \left\lfloor \tfrac{\lambda_{kl}}{d_l} \right\rceil \right)
	+ \T(q b_l) + (l-1) \T(q \lambda_{lj}) + \T(q d_l) = \\
	&= O\left( m^2 \mul(m^2 D) + n m \mul(n^2 H, D) \right).
\end{align*}

Now consider the swap operation. It performs the following calculations: \\
\begin{algorithm}[H]
$b_k \leftrightarrow b_{k-1}$\;
\For{$j := 1$ \KwTo $k-2$}{
	$\lambda_{k,j} \leftrightarrow \lambda_{k-1,j}$\;
}
$d_{k-1}' := \frac{d_{k-2} d_k + \lambda_{k,k-1}^2}{d_{k-1}}$\;
\For{$i := k+1$ \KwTo $n$}{
	$\lambda_{i,k}' := \frac{d_k \lambda_{i,k-1} - \lambda_{k,k-1} \lambda_{i,k}}{d_{k-1}}$\;
	$\lambda_{i,k-1} := \frac{d_{k-1}' \lambda_{i,k} + \lambda_{k,k-1} \lambda_{i,k}'}{d_k}$\;
	$\lambda_{i,k} := \lambda_{i,k}'$\;
}
$d_{k-1} := d_{k-1}'$\;
\end{algorithm}

The major operations have a similar structure than the recursive formula of the Bareiss algorithm (\ref{bareiss-rec}), so a very similar calculation can be performed. The differences are that $B$ is replaced by $D$ (but they both dominate $m F$), that all operations are performed $O(n)$ times except the inversion of the denominator, which is done twice. This leads to the total time of the swap operation, which is:
\begin{equation*}
	\T(\text{swap}) = O\left( m^3 \mul(m D) + n m^3 \mul(D) \right).
\end{equation*}

The third main part is the comparison in the main \textbf{if} statement, which can be expressed equivalently as:
\begin{equation*}
	d_{k-2} d_k + \lambda_{k,k-1}^2 < \delta d_{k-1}^2.
\end{equation*}
The multiplications, like in the swap part, take $O(m^2 \mul(D))$ time, and the comparison itself is $O(m^2 \mul(m D))$ by (\ref{cmp-time}), so
\begin{equation*}
	\T(\text{cmp}) = O(m^2 \mul(m D)).
\end{equation*}

Now we can put together the running time of the whole algorithm. It has a main \textbf{while} loop, where each iteration is either a swap step or a reduction step (not to be confused with $\T(\text{swap})$ and $\T(\text{red})$). The swap step makes a reduction, a comparison and a swap, and the reduction step makes a comparison and $k-1$ reductions. Their running time is:
\begin{align*}
	\T(\text{swap step})
	&= \T(\text{cmp}) + \T(\text{red}) + \T(\text{swap})
	= O\left( m^2 \mul(m^2 D) + n m^3 \mul(D) \right), \\
	\T(\text{red step})
	&= \T(\text{cmp}) + (k-1) \T(\text{red})
	= O\left( n m^2 \mul(m^2 D) + n^2 m \mul(n^2 H, D) \right).
\end{align*}
If $N$ is the number of main iterations, we proved in Lemma~\ref{lem:iter-bound} (combined with Lemma~\ref{lem:H-bound}) that $N = O\left( n^3 H K_\delta \right)$. Therefore, the running time of the LLL algorithm is:
\begin{align*}
	\T_{\mathbb{Z}[\theta]}(\text{LLL})
	&\leq N \T(\text{swap step}) + N \T(\text{red step}) = \\
	&= N O\left( n m^2 \mul(m^2 D) + n^2 m \mul(n^2 H, D) \right) = \\
	&= O\left( n^4 m H K_\delta (m \mul(n^6 m^2 H^2 K_\delta) + n^5 H K_\delta \mul(n^2 H)) \right),
\end{align*}
where, again, the meaning of the variables are the following:
\begin{itemize}
	\item $H = O\left( m A + m^2 F + m \log n \right)$ by Lemma~\ref{lem:H-bound},
	\item $n$ is the dimension of the lattice,
	\item $m$ is the degree of the algebraic number field,
	\item $F = \log\left( \max_{i=0}^{m-1} |f_i| + 1 \right)$, where $f(x) = x^m + \sum_{i=0}^{m-1} f_i x^i$ is the minimal polynomial of the primitive element $\theta$ in the number field,
	\item $A = \max_{i=1}^{n} \s(b_i)$, the coefficient size of the input vectors,
	\item $K_\delta = \frac{1}{\log \frac{1}{\delta}}$, where $\delta$ is the parameter of the LLL algorithm between $1/4 < \delta < 1$.
\end{itemize}

For comparison, the running time for integers is the following (see e.g.\ in \cite{lll}):
\begin{equation*}
	\T_\mathbb{Z}(\text{LLL})
	= O\left( n^4 \log B \mul(n \log B) K_\delta \right),
\end{equation*}
where $B = \max_{i=1}^{n} \lVert b_i \rVert_2^2$.

In the following table, we compare the results for several different $\mul(\cdot)$ functions in both $\mathbb{Z}[\theta]$ and $\mathbb{Z}$. For better comparison, we define $A := \log \max_{i=1}^{n} \lVert b_i \rVert_\infty$ for $\mathbb{Z}$, so we have $\log B \leq 2 A + \log n$.

\begin{center}
\begin{tabular}{|l|l|l|}
	\hline
	$\mul(X)$
	& $\T_{\mathbb{Z}[\theta]}(\text{LLL})$
	& $\T_\mathbb{Z}(\text{LLL})$
	\\ \hline
	$X^2$
	& $O\left( n^{16} m^{11} (A + m F + \log n)^5 K_\delta^3 \right)$
	& $O\left( n^6 (A + \log n)^3 K_\delta \right)$
	\\ \hline
	$X^{\log_2 3}$
	& $O\left( n^{13.6} m^{9.4} (A + m F + \log n)^{4.2} K_\delta^{2.6} \right)$
	& $O\left( n^{5.6} (A + \log n)^{2.6} K_\delta \right)$
	\\ \hline
	$X \log X \log \log X$
	& $\tilde{O}\left( n^{10} (n + m^3) m^4 (A + m F + \log n)^3 K_\delta^2 \right)$
	& $\tilde{O}\left( n^5 (A + \log n)^2 K_\delta \right)$
	\\ \hline
\end{tabular}
\end{center}

In the last row, $\tilde{O}(N) = O(N \log N \log \log N)$ as in Section~\ref{sec:bareiss}.

\subsection{Notes on the LLL result} \label{sec:notes}

We proved that the LLL algorithm does not suffer from exponential coefficient growth even for exact algebraic numbers, and it has polynomial time complexity. However, its running time is significantly different from the integer version, not only in the presence of additional parameters ($m$ and $F$), but also in the order of the basic parameters ($n$ and $A$). This shows how much harder it is to contain the coefficient size ($\s(\cdot)$) than the normal size ($|\cdot|$), for example while $\lVert b_i \rVert_2^2 \leq n B$ for most $i$, we have $\s(b_i) = O(n^5 H^2 K_\delta)$.

However, our actual result for $\mathbb{Z}[\theta]$ is just a very pessimistic upper bound for the worst-case complexity. We strongly believe that the algorithm is much faster in practice. For example, the number of iterations in the algorithm is $N = O\left( n^3 H K_\delta \right)$, but this is only a theoretical limit, and in practice, it can often be just a few (i.e.\ $O(n)$) steps. Furthermore, we used $L_0$, the size of the shortest vector in the lattice, and we calculated a worst-case theoretical lower bound for it: $\log \frac{1}{L_0} = O\left( n m A + n m^2 F + n m \log n \right)$. But in practice, there is no special reason why the shortest vector would be so extremely small. If we can make an assumption that it is constant (i.e.\ $O(1)$), then the running time can be reduced by several powers. It is easy to check that e.g.\ for basic multiplication ($\mul(X) = X^2$), these two practical assumptions reduce $n^{16}$ to $n^{10}$.

We suspect that the powers can be reduced even further in average. It is out of scope of the present theoretical article but is a subject of future research to perform systematic measurements on the actual running time to confirm these claims.

\end{document}